\documentclass{ws-procs10x7}
\usepackage{balance}
\usepackage{epstopdf}
\newcolumntype{d}[1]{D{.}{.}{#1}}

\makeindex
\begin{document}

\title{Inclusive $B \to X_c l \nu$ decay spectra at Belle and the determination of $|V_{cb}|$}

\author{P. Urquijo\\
(For the Belle Collaboration)}

\address{School of Physics, University of Melbourne, Parkville, Victoria 3010, Australia\\$^*$E-mail: phill@physics.unimelb.edu.au}


\twocolumn[\maketitle\abstract{We report measurements of the inclusive electron energy spectrum and hadron invariant mass spectrum for charmed semileptonic decays of $B$ mesons in a 140 fb$^{-1}$ data sample collected with the Belle detector at the KEKB $e^+e^-$ collider.  We determine the first four central moments and partial branching fraction of the electron energy spectrum for electron energy thresholds from 0.4 to 2.0 GeV, and the first two central and second non-central moments of the hadron invariant mass spectrum for lepton energy thresholds from 0.7 to 1.9 GeV. Using these measurements and Belle measurements of the photon energy moments in $B\to X_s\gamma$ decays, we determine the CKM matrix element $|V_{cb}|$, the $b$ quark mass and higher order non-perturbative parameters that appear in the Heavy Quark Expansion by performing a global fit analysis in the kinetic mass and 1S schemes.}
\keywords{B physics; CKM matrix; Semileptonic decays; Heavy quarks.}
]

\section{Introduction}
The most precise determinations of the Cabibbo-Kobayashi-Maskawa (CKM)
matrix element~$|V_{cb}|$~\cite{Kobayashi:1973fv} are obtained using
combined fits to inclusive $B$~decay
distributions~\cite{Bauer:2004ve,Buchmuller:2005zv}. These determinations
are based on calculations of the semileptonic decay rate in the
frameworks of the Operator Product Expansion (OPE)~\cite{wilson} and Heavy Quark
Effective Theory (HQET)~\cite{Bauer:2004ve,Benson:2003kp} which
predict this quantity in terms of $|V_{cb}|$, the $b$-quark mass
$m_b$, and non-perturbative matrix elements that enter at the
order $1/m^2_b$.
The spectator model decay rate is the leading term in a well-defined 
expansion controlled by  the parameter $\Lambda _{\rm QCD}/m_b$~\cite{Benson:2003kp,gremm-kap,falk,Gambino:2004qm} with
non-perturbative corrections arising to order
 $1/m_b^2$. The key issue in this approach is the ability to separate non-perturbative 
 corrections and perturbative 
 corrections (expressed in powers of  $\alpha _s$).  
High precision comparison of theory and experiment requires a precise determination of the heavy quark masses, as well as the non-perturbative matrix elements that enter the expansion.  We make use of the HQEs \cite{Bauer:2004ve,Gambino:2004qm,Benson:2004sg} that express the semileptonic decay width  $\Gamma_{SL}$, moments of
the lepton energy and hadron mass spectra in $B\to
X_c\ell\nu$~decays and the photon energy spectrum in $B\to
X_s\gamma$~decays in terms of the running kinetic quark masses $m_b^{\rm kin}$ and $m_c^{\rm kin}$ as well as the 1S $b$-quark mass $m_b^{\rm 1S}$. These schemes should ultimately yield consistent results for $|V_{cb}|$.  The precision of the $b$-quark mass is also important for $|V_{ub}|$, a limiting factor in the uncertainty on the unitarity triangle.

The present results are based on a $140\,{\rm fb}^{-1}$ data sample collected at the $\Upsilon (4S)$ resonance with the Belle detector
at the KEKB~\cite{KEKB} asymmetric energy $e^+ e^-$ collider, containing $1.52 \times 10^8$ $B \overline B$ pairs.  An additional $15\,{\rm fb}^{-1}$ data sample taken at 60MeV below the $\Upsilon (4S)$ resonance is used to perform subtraction of background from the continuum $e^+e^- \rightarrow q \bar q$ process.

\section{Inclusive spectral moments}

We first identify hadronic events~\cite{hadron}, then fully reconstruct one $B$ meson in one of several hadronic modes to determine its charge, flavour, and momentum ($B_{\rm tag}$).
The number of $B^+$  and $B^0$  candidates in the  signal region,  after background subtraction, is $63185\pm621$(stat.) and $39504\pm392$(stat.), respectively.  We search for leptons produced by semileptonic $B$ decays on the non-tag side.   We partially recover the effect of bremsstrahlung
by searching for a photon  around the electron direction.
The reconstructed lepton momentum  spectrum is contaminated by background processes, evaluated and subtracted from the distribution before the extraction of the moments.  Backgrounds are predominantly continuum, combinatorial,
 cascade charm decays $b \rightarrow c \rightarrow q \ell \nu$,  $J/ \psi$, $\psi (2S)$, Dalitz decays, photon conversions,  fake leptons and  $B \to X_u \ell \nu$ decays. We use the LLSW~model~\cite{LLSW} to predict the relative abundance
and form factor shapes of the different components in $B\to D^{**}\ell\nu$. 

\subsection{Electron Energy Spectrum}
\begin{figure}
\includegraphics[width=0.24\textwidth]{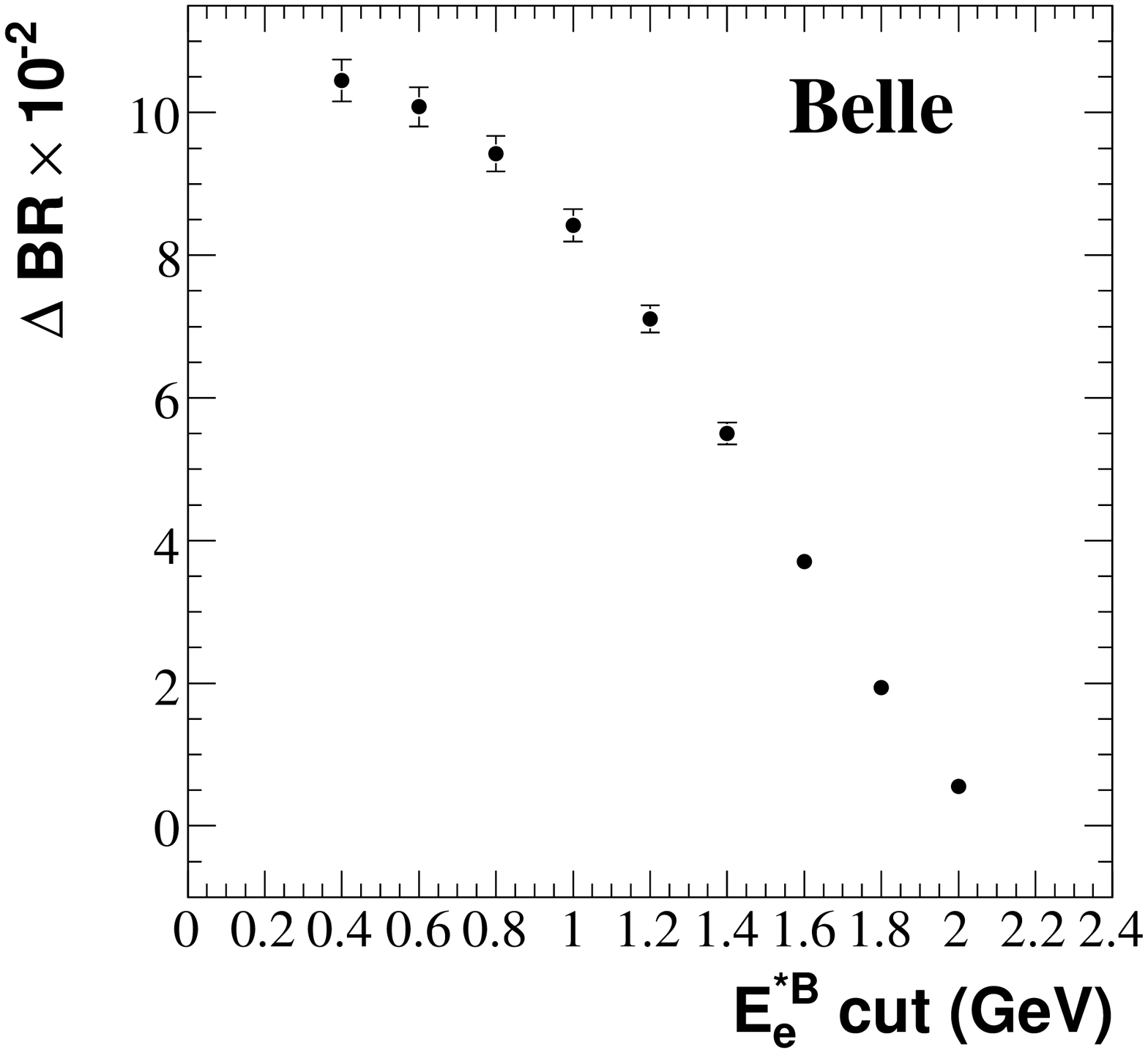}
\includegraphics[width=0.24\textwidth]{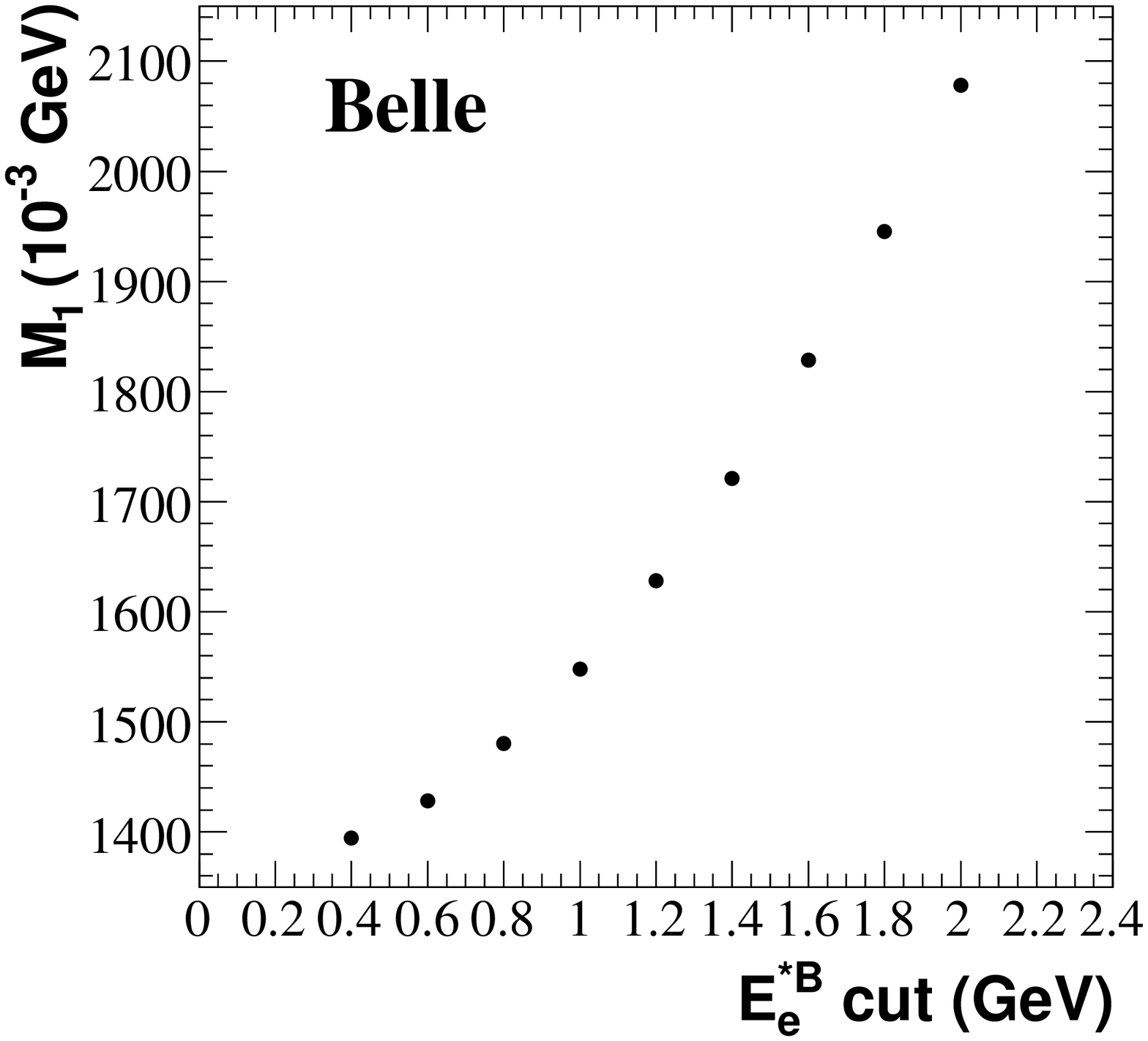}\\
\includegraphics[width=0.24\textwidth]{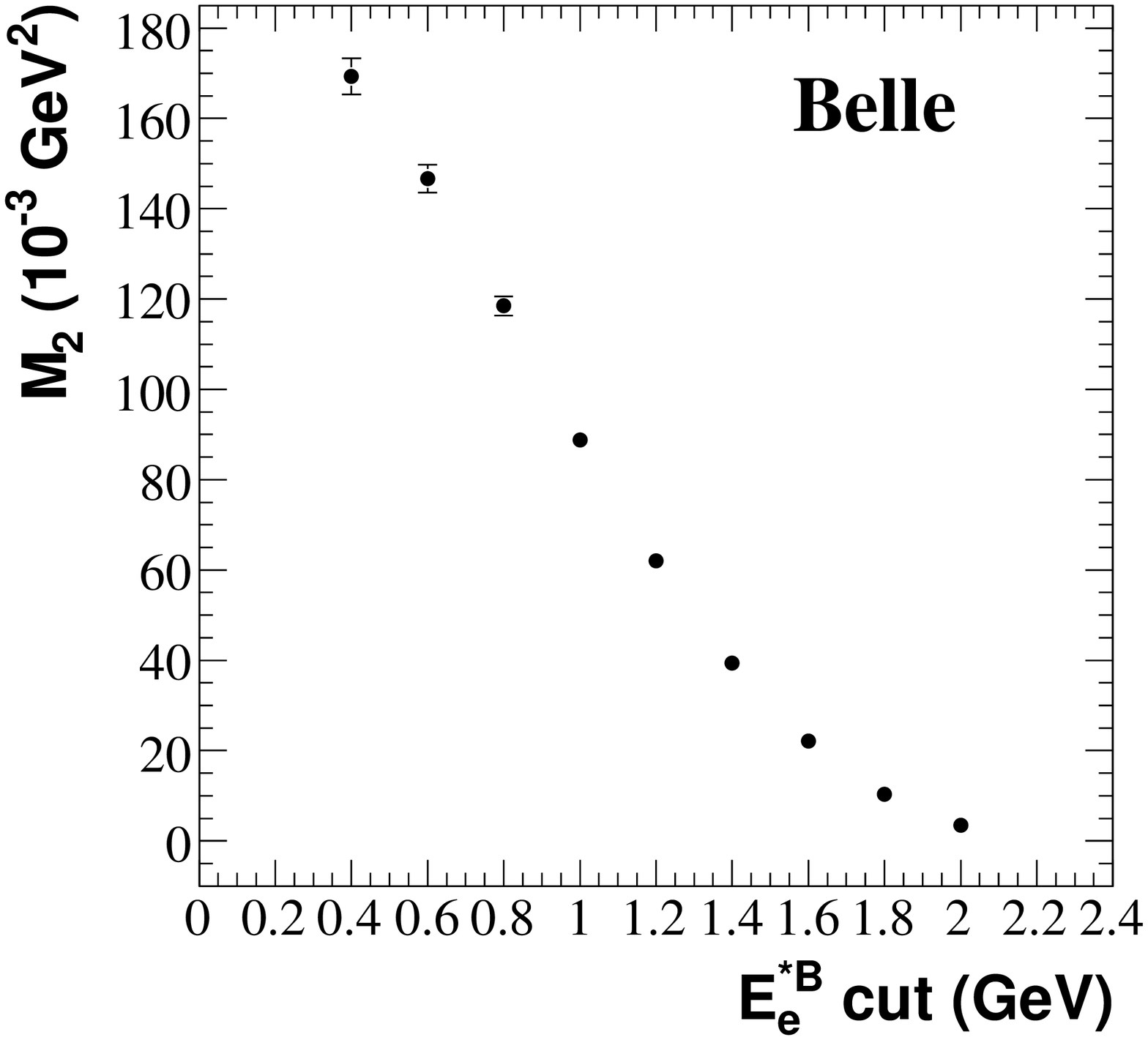}
\includegraphics[width=0.24\textwidth]{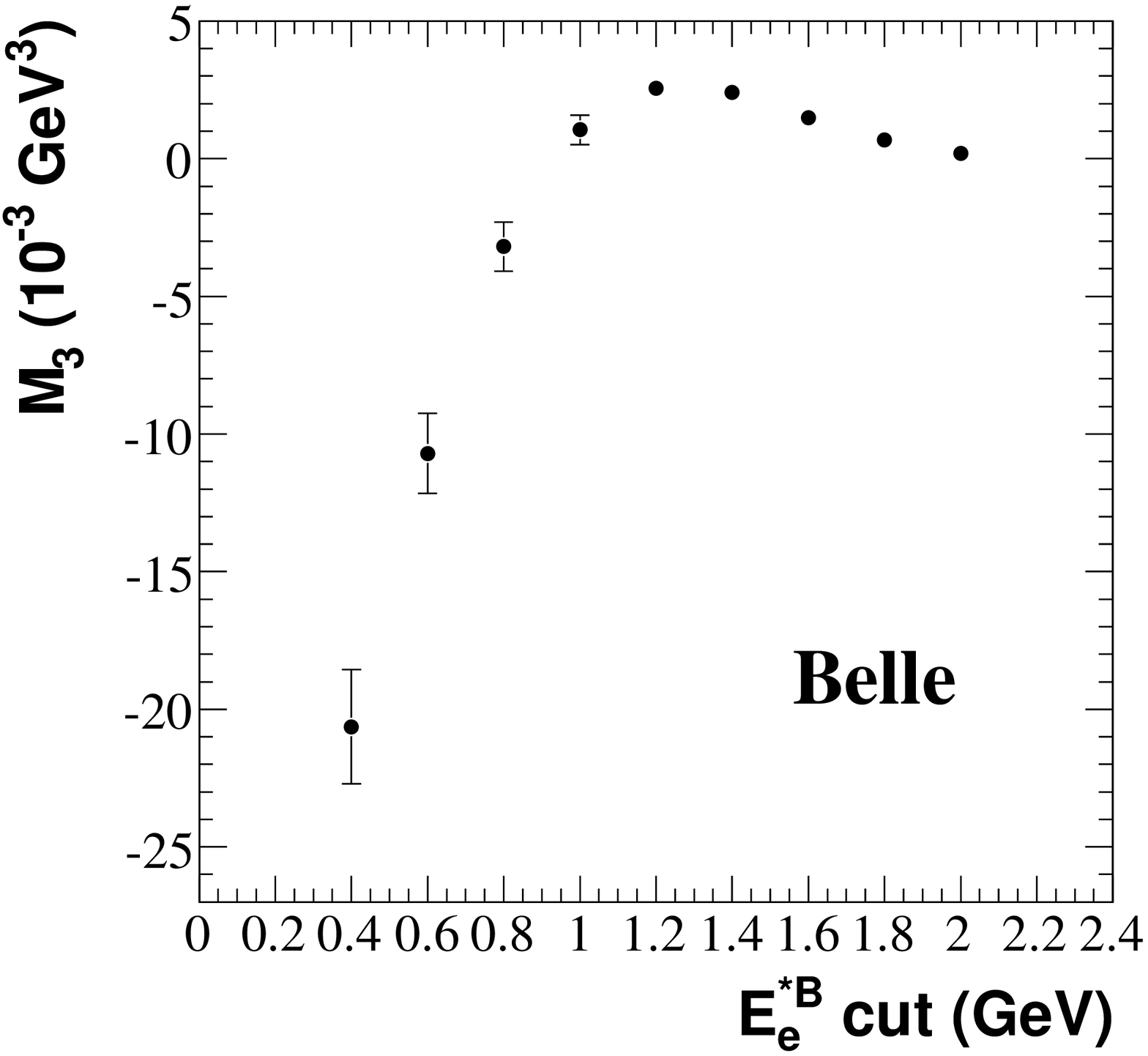}\\
\includegraphics[width=0.24\textwidth]{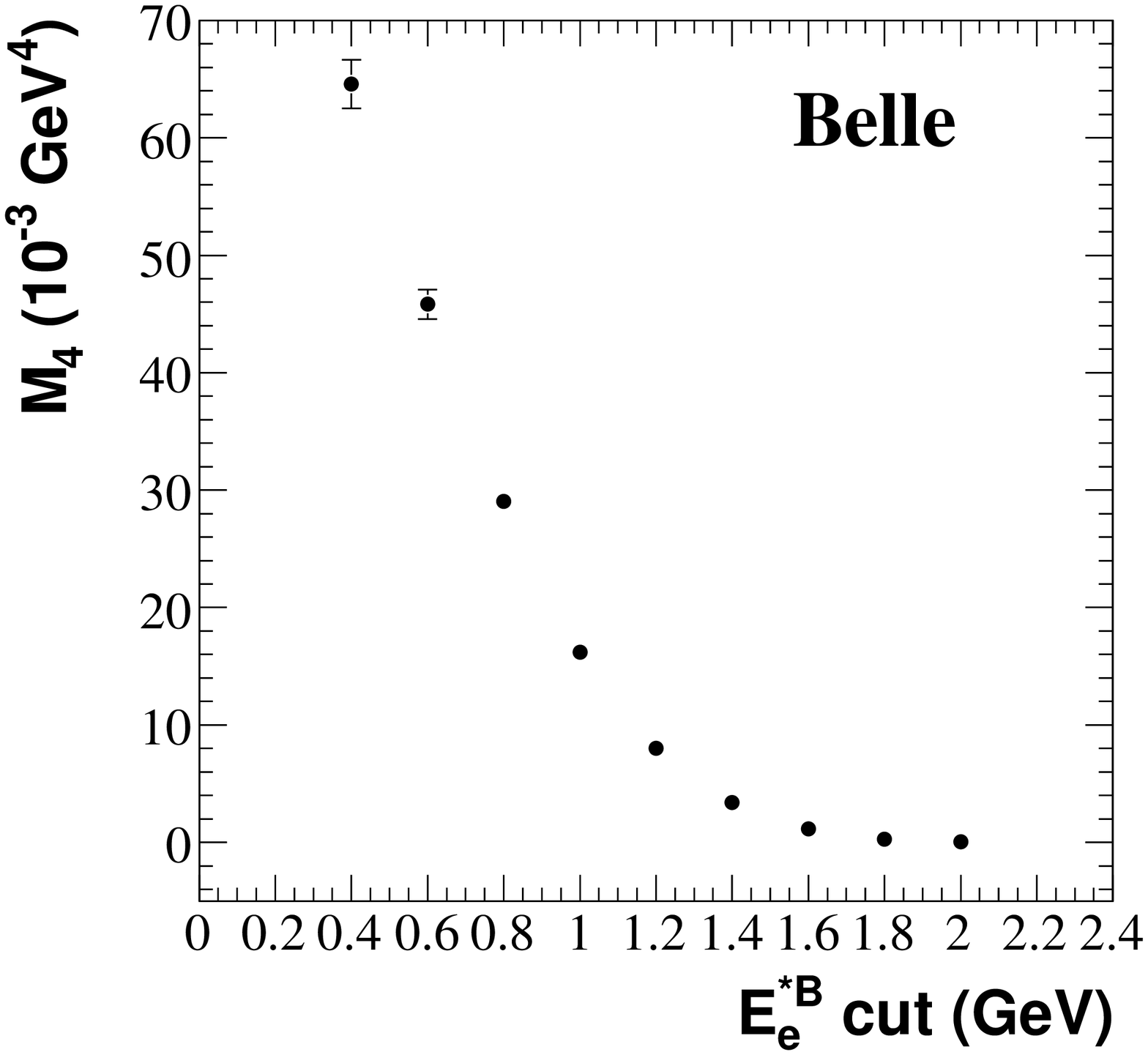}
\caption{Moments and partial branching fractions as a function of the electron energy cut-off, $E_{cut}$}
\label{fig1}
\end{figure}

To measure the moments of the electron energy spectrum, we determine the true electron energy spectrum by unfolding~\cite{ref:13} the measured spectrum for distortions by various detector effects, in the $B$ meson rest frame, $E_{e}^{*B}$. 
 The unfolded spectrum is corrected for QED radiative effects using PHOTOS~\cite{PHOTOS}.
Belle measures the the  $B^0$ and $B^+$ weighted average partial branching fractions~$\mathcal{B}(B\to
X_c\ell\nu)_{E_\ell>E_\mathrm{min}}$ and the first four moments of the electron energy spectrum in $B\to
X_ce\nu$, $\langle E_\ell\rangle_{E_\ell>E_\mathrm{min}}$, $\langle
(E_\ell-\langle E_\ell\rangle)^2\rangle_{E_\ell>E_\mathrm{min}}$,
$\langle (E_\ell-\langle
E_\ell\rangle)^3\rangle_{E_\ell>E_\mathrm{min}}$ and $\langle
(E_\ell-\langle E_\ell\rangle)^4\rangle_{E_\ell>E_\mathrm{min}}$, for
 electron energy thresholds, $E_\mathrm{min}$, from 0.4 to 2.0~GeV)~\cite{el} (Fig. \ref{fig1}) .   
The principal systematic errors originate from event selection, electron identification, background estimation and signal model dependence.  
The independent partial branching fractions at $E_{\rm cut}=$0.6 GeV are $\Delta \mathcal{B}(B^+ \to X_c e \nu$)=$(10.34 \pm 0.23({\rm stat.}) \pm 0.25({\rm sys.}))$\% and $\Delta \mathcal{B}(B^0 \to X_c e \nu$)=$(9.80 \pm 0.29({\rm stat.}) \pm 0.21(\rm{ sys.}))$\%; consistent with our previous measurements~\cite{okabe}.  The $\Delta \mathcal{B} (B^+ \to X_c e \nu)$/$\Delta \mathcal{B} (B^0 \to X_c e \nu)$ ratio, at $E_{\rm cut}=$0.4 GeV, is $1.07 \pm 0.04({\rm stat.}) \pm 0.03({\rm sys.})$, consistent with the $B^+/B^0$ lifetime ratio~\cite{PDG}. 

\subsection{Hadron Mass Spectrum}
\begin{figure}
  \begin{center}
    \includegraphics[width=0.48\textwidth]{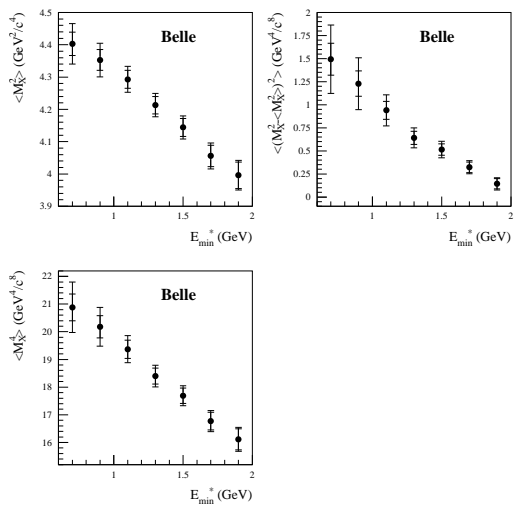}
  \end{center}
  \caption{The hadron invariant mass moments as a function of $E_{\rm min}$. The error bars indicate the statistical and
   total experimental errors.} \label{fig:2}
\end{figure}

The 4-momentum $p_X$ of the hadronic system~$X$ recoiling against
$\ell\nu$ is determined by summing the 4-momenta of the remaining
charged tracks and unmatched clusters. Events with particles missing in the reconstruction are rejected by
requiring $|M^2_\mathrm{miss}|<3$~GeV$^2$/$c^4$.  To improve the resolution in $M^2_X$, we constrain the neutrino mass
to zero, and recalculate the 4-momentum of the $X$~system,
$p'_X = (p_\mathrm{LER}+p_\mathrm{HER})-p_{B_{tag}}-p_\ell-p_\nu$, where LER and HER refer to the low energy and high energy beams,
respectively.
We measure the $M^2_X$~spectrum in from 0 to
15~GeV$^2$/$c^4$,  and unfold the finite detector resolution in this
distribution to a range of $M^2_D$ to
about 15 GeV$^2$. The bin width is 1~GeV$^2$, except around
the narrow states, ($D$, $D^*$, $D_1$ and $D^*_2$) where smaller
bin sizes are chosen.  Belle measures the first, second central and second non-central
moments of the unfolded $M^2_X$~spectrum in $B\to X_c\ell\nu$, $\langle
M^2_X\rangle_{E_\ell>E_\mathrm{min}}$, $\langle(M^2_X-\langle
M^2_X\rangle)^2\rangle_{E_\ell>E_\mathrm{min}}$ and $\langle
M^4_X\rangle_{E_\ell>E_\mathrm{min}}$ for lepton
energy thresholds, $E_\mathrm{min}$, from 0.7 to 1.9 GeV~\cite{mx} (Fig.~\ref{fig:2}).
Principal systematic errors originate from background estimation, unfolding and signal model dependence.

\section{HQE parameters}
Using these measurements and Belle measurements of the photon energy moments in $B\to X_s\gamma$ decays~\cite{Abe:2005cv}, we determine the CKM matrix element $|V_{cb}|$, and HQE parameters by performing global fit analyses in the kinetic and 1S schemes~\cite{bellefit}.  We exclude
measurements that do not have corresponding theoretical
predictions and those with high cutoff energies ({\it i.e.} semileptonic moments with
$E_\mathrm{min}>1.5$~GeV and  photon energy moments with
$E_\mathrm{min}>2$~GeV).  All fit results are preliminary.

\subsection{1S Fit}
\begin{figure}
  \includegraphics[width=0.24\textwidth]{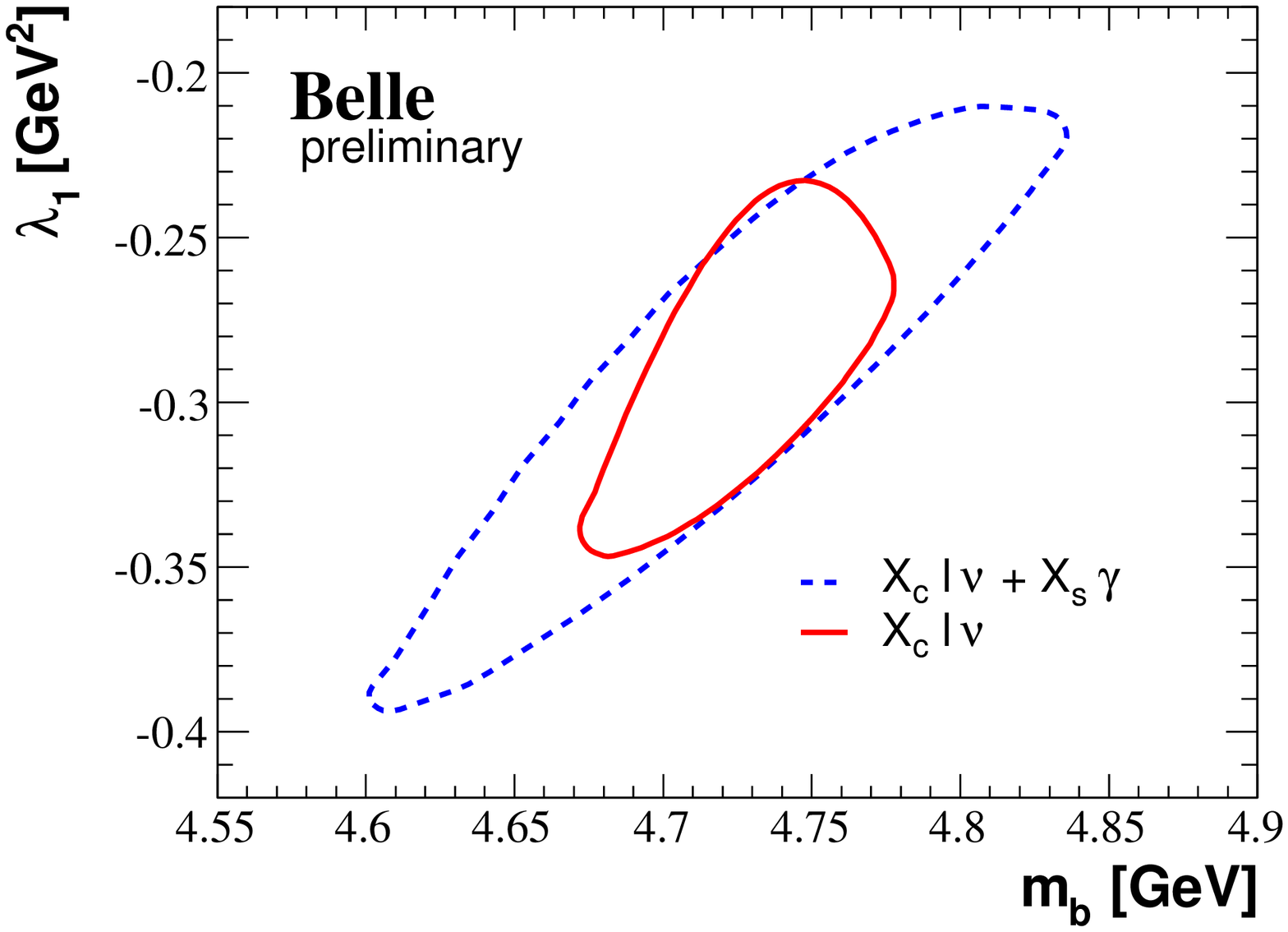}
  \includegraphics[width=0.24\textwidth]{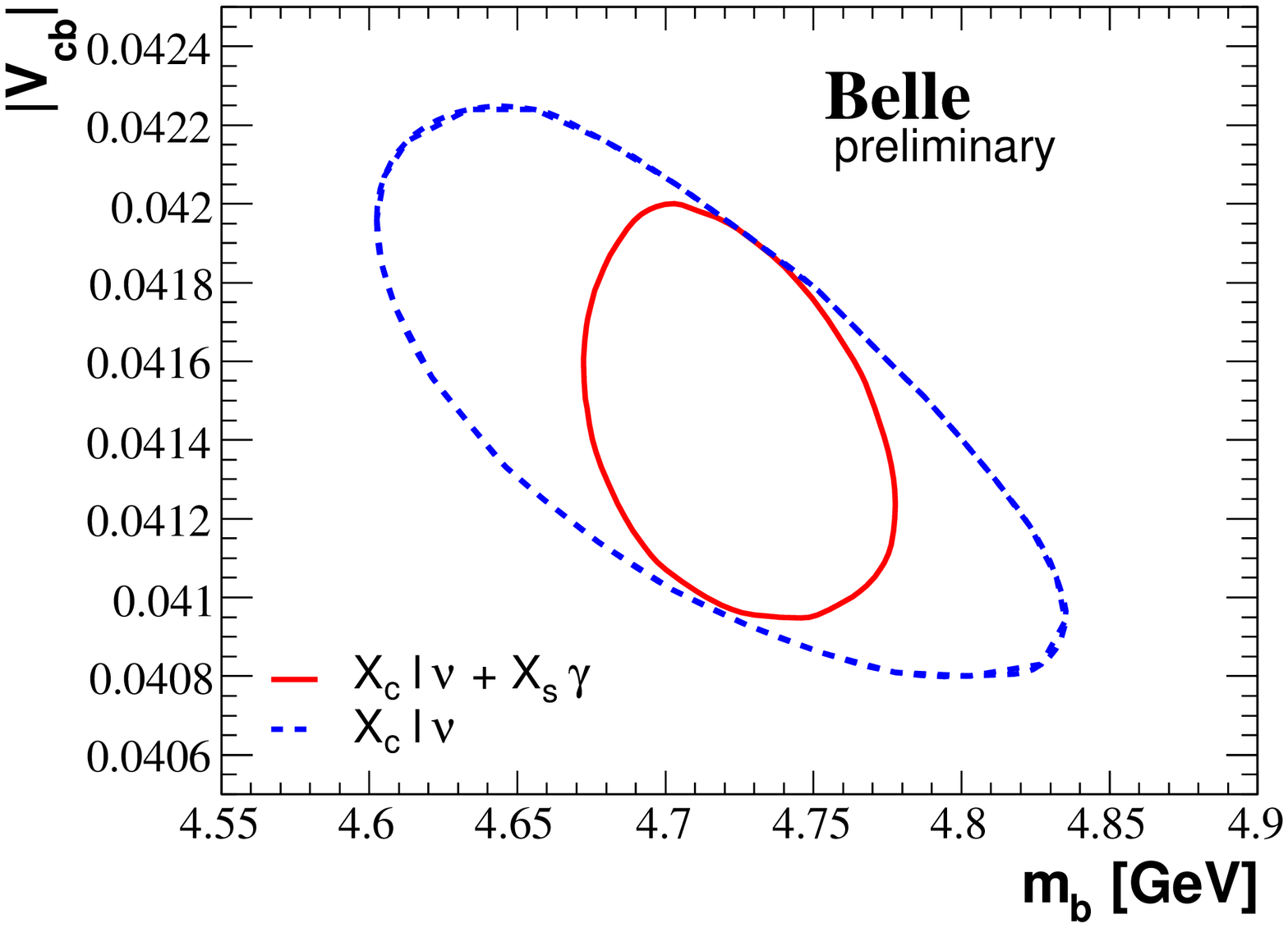}
  \caption{Fit results for $m_b^\mathrm{1S}$ and $\lambda_1$ (left), and
    $m_b^\mathrm{1S}$ and $|V_{cb}|$ (right) to 
    $B\to X_c\ell\nu$~data only   (dashed blue line) and $B\to X_c\ell\nu$ and $B\to
    X_s\gamma$~data combined (solid
    red line). The ellipses are $\Delta\chi^2=1$.}
    \label{fig:3}
\end{figure}

The inclusive spectral moments of $B\to X_c\ell\nu$ decays have been
derived in the 1S scheme up to
$\mathcal{O}(1/m_b^3)$~\cite{Bauer:2004ve}. The theoretical
expressions for the truncated moments are given in terms of
HQE parameters, and coefficients determined by theory, which are functions of
$E_\mathrm{min}$. The non-perturbative corrections are parametrized by
$\Lambda$ ($\mathcal{O}(m_b)$), $\lambda_1$ and $\lambda_2$
($\mathcal{O}(1/m_b^2)$), and $\tau_1$, $\tau_2$, $\tau_3$, $\tau_4$,
$\rho_1$ and $\rho_2$ ($\mathcal{O}(1/m_b^3)$).  We find the following results for the fit
parameters (Fig. \ref{fig:3}),
\begin{eqnarray*}
  |V_{cb}| & = & (41. 49\pm 0.52_\mathrm{fit}\pm 0.20_{\tau_B})\times
   10^{-3}~,\\
   m_b^\mathrm{1S} & = & (4.729\pm 0.048)~\mathrm{GeV}~,{\rm~and}\\
   \lambda_1 & = & (-0.30\pm 0.04)~\mathrm{GeV}^2~.
\end{eqnarray*}
The first error is from the fit including experimental and
theory errors, and the second error (on $|V_{cb}|$ only) is due to the
uncertainty on the average $B$~lifetime ($\chi^2/$n.d.f.$=5.7/17$).  Using the
partial branching fraction measurement at
$E_\mathrm{min}=0.6$~GeV, we obtain for the full semileptonic branching
ratio, $ {\mathcal B}(B\to X_c\ell\nu)=(10.62\pm 0.25)\%~$.

\subsection{Kinetic fit}
\begin{figure}
  \begin{center}
    \includegraphics[width=0.48\textwidth]{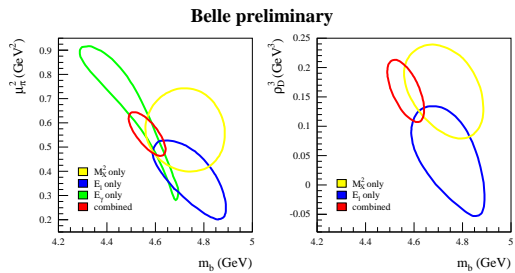}
  \end{center}
  \caption{The kinetic scheme fit repeated
    using lepton energy moments only, hadron mass moments only and
    photon energy moments only. The ellipses are $\Delta\chi^2=1$.}
    \label{fig:6}
\end{figure}

Spectral moments of $B\to X_c\ell\nu$~decays have been derived up to
$\mathcal{O}(1/m^3_b)$ in the kinetic
scheme~\cite{Gambino:2004qm}. The
theoretical expressions used in the fit contain improved
calculations of the perturbative corrections to the lepton energy
moments~\cite{ref:2} and account for the $E_\mathrm{min}$~dependence
of the perturbative corrections to the hadronic mass
moments~\cite{Uraltsev:2004in}. For the $B\to X_s\gamma$~moments, the
(biased) OPE prediction and the bias correction have been
calculated~\cite{Benson:2004sg}.  All these expressions depend on the $b$- and $c$-quark masses
$m_b(\mu)$ and $m_c(\mu)$, the non-perturbative parameters $\mu^2_\pi(\mu)$ and $\mu^2_G(\mu)$ ($\mathcal{O}(1/m^2_b)$),
$\tilde\rho^3_D(\mu)$ and $\rho^3_{LS}(\mu)$ ($\mathcal{O}(1/m^3_b)$),
and $\alpha_s$~\cite{ref:3}. The CKM element $|V_{cb}|$ is a
free parameter in the fit, related to the semileptonic
width $\Gamma(B\to X_c\ell\nu)$~\cite{Benson:2003kp}.
We find the following results for the fit parameters (Fig. \ref{fig:6}),
\begin{eqnarray*}
|V_{cb}|&=&(41.93\pm 0.65\pm 0.48
\pm0.63)\times 10^{-3},~\\
\mathcal{B}_{X_c\ell\nu}&=& (10.590 \pm 0.164\pm  0.006)\%~,\\
m_b  &=& (4.564 \pm 0.076 \pm 0.003)~{\rm GeV}~, \\
m_c &=& (1.105 \pm 0.116 \pm 0.005)~ {\rm GeV}~{\rm and}\\
 \mu^2_\pi&=& (0.557 \pm 0.091 \pm 0.013)~{\rm GeV}^2. 
\end{eqnarray*}
The first error is from the fit 
(experimental error, non-perturbative 
and bias corrections). The second error is 
obtained by varying $\alpha_s$ in the expressions for the moments. The last error is a 1.5\% uncertainty from
the theoretical expression for the semileptonic width~\cite{Benson:2003kp} ($\chi^2/$n.d.f.$=17.8/24$). 

\section*{Acknowledgments}
We thank our colleagues at Belle; C.~Schwanda, E.~Barberio, and A.~Limosani for providing input to this proceeding.
\balance


\begin{thebibliography}{9}
\bibitem{ref:3}
  Non-perturbative parameters in the kinetic
  scheme are defined at the scale $\mu=1$~GeV.
\end{thebibliography}
\end{document}